\documentclass[11pt]{article}

\usepackage[T1]{fontenc}
\usepackage[english]{babel}
\usepackage{amsmath,amssymb,amsthm,thmtools}
\usepackage{fancyhdr}
\usepackage{setspace}
\usepackage{graphicx}
\usepackage{colortbl}
\usepackage{pgf}
\usepackage{listings}
\usepackage{indentfirst}
\usepackage{algorithm}
\usepackage[noend]{algpseudocode}
\usepackage{optidef}
\usepackage{textcomp, gensymb}
\usepackage{tikz}
\usepackage{subcaption}
\usepackage{mathtools}
\usepackage{pythonhighlight}
\usepackage{svg}
\usepackage{verbatim}
\usepackage{graphicx} 
\usepackage{setspace} 
\usepackage{xspace}
\usepackage{tabto}
\usepackage{array}
\usepackage{dirtytalk}
\usepackage{todonotes}
\usepackage{tabularray}
\usepackage{tabularx}
\usepackage{enumerate}
\usepackage{makecell}
\usepackage{xcolor}
\usepackage{wrapfig}
\usepackage[makeroom]{cancel}
\usepackage[noEnd=true]{algpseudocodex}
\usepackage{scrextend}
\usepackage{changepage}
\usepackage{multicol}
\usepackage{comment}
\usepackage{enumitem}

\DeclarePairedDelimiter\floor{\lfloor}{\rfloor}

\makeatletter
\renewcommand{\@biblabel}[1]{#1.}
\makeatother

\usepackage[margin=2.5cm]{geometry}

\usepackage[]{color-edits}
\addauthor{UF}{red}
\newcommand{\ufc}[1]{{\UFcomment{#1}}}
\newcommand{\ufe}[1]{{\UFedit{#1}}}

\addauthor{IP}{blue}

\usepackage{pgfplots}
\pgfplotsset{compat=1.15}
\usepackage{mathrsfs}
\usetikzlibrary{arrows}

\usepackage[colorlinks, linkcolor=blue, urlcolor=blue]{hyperref}
\hypersetup{
    colorlinks,
    linkcolor={red!50!black},
    citecolor={blue!50!black},
    urlcolor={blue!80!black}
}
\usepackage[noabbrev,capitalize]{cleveref}

\newtheorem{theorem}{Theorem}
\newtheorem{lemma}[theorem]{Lemma}
\newtheorem{observation}[theorem]{Observation}
\newtheorem{definition}{Definition}
\newtheorem{conjecture}{Conjecture}
\newtheorem{proposition}{Proposition}
\newtheorem{corollary}{Corollary}
\newtheorem{claim}[theorem]{Claim}

\newtheorem{remark}[theorem]{Remark}

\begin{document}

\title{Large cliques and large independent sets: can they coexist?}

\author{Uriel Feige\thanks{Weizmann Institute of Science, Israel. {\tt uriel.feige@weizmann.ac.il}} \and Ilia Pauzner\thanks{Weizmann Institute of Science, Israel. {\tt ilpauzner@gmail.com}}}

\maketitle

\begin{abstract}
    For a graph $G$ and a parameter $k$, we call a vertex {\em $k$-enabling} if it belongs both to a clique of size $k$ and to an independent set of size $k$, and we call it {\em $k$-excluding} otherwise. Motivated by issues that arise in secret sharing schemes, we study the complexity of detecting vertices that are $k$-excluding. We show that for every $\epsilon$, for sufficiently large $n$, if $k > (\frac{1}{4} + \epsilon)n$, then every graph on $n$ vertices must have a $k$-excluding vertex, and moreover, such a vertex can be found in polynomial time. In contrast, if $k < (\frac{1}{4} - \epsilon)n$, a regime in which it might be that all vertices are $k$-enabling, deciding whether a graph has no $k$-excluding vertex is NP-hard.
\end{abstract}

\section{Introduction}

A clique 
in a graph $G(V,E)$ is a set $S$ of vertices such that every pair of vertices in the set share an edge ($(u,v) \in E)$ for every $u,v\in S$). An independent set is a set of vertices such that no pair of vertices in the set share an edge ($(u,v) \not\in E)$ for every $u,v\in S$).

For a graph $G$ and a parameter $k$, we call a vertex {\em $k$-enabling} if it belongs both to a clique of size $k$ and to an independent set of size $k$, and we call it {\em $k$-excluding} otherwise. We refer to a graph in which all vertices are $k$-enabling as a $k$-enabling graph. We define the function $k(n)$ whose value for every integer $n$ is equal to the largest integer $k$ such that there is a graph on $n$ vertices that is $k$-enabling. One theme of this work is existential, to provide lower bounds and upper bounds on the function $k(n)$.

Trivially, the function $k(n)$ is monotonically non-decreasing ($k(n+1) \ge k(n)$) and has a Lipschitz constant of~1 ($k(n+1) \le k(n) + 1$).

\begin{proposition}
    The function $k(n)$ is monotonically non-decreasing and has a Lipschitz constant of~1. 
\end{proposition}

\begin{proof}
    {\bf Monotonicity.} Given an $n$ vertex $k$-enabling graph $G$, we can create from it a graph on $n+1$ vertices that is $k$-enabling by replacing any vertex $v$ by two vertices, $v_1$ and $v_2$, each with exactly the same set of neighbors as $v$. (The resulting graph is $k$-enabling regardless of whether $(v_1,v_2)$ is an edge or not.) 
    
    {\bf Lipschitz.} Given an $n+1$ vertex $k$-enabling graph $G$, we can create from it a graph on $n$ vertices that is $(k-1)$-enabling by dropping any vertex from $G$.  
\end{proof}

There is a simple lower bound on function $k(n)$. 

\begin{proposition}
    \label{pro:lowerBound}
    For every $n$ divisible by~4, $k(n) \ge \frac{n}{4} + 1$.
\end{proposition}

\begin{proof}
    We construct an infinite family of graphs, one graph for every positive integer $d$. We refer to graphs in this family as $4P_d$ (the name $4P$ is meant to represent the fact that these graphs are derived from paths on 4 vertices). For every $d$, the graph $4P_d$ has $4d$ vertices, and is $(d+1)$-enabling.

    For $d=1$, the graph $4P_1$ is simply a path on 4 vertices. We refer to the two end-vertices of the path (that have degree~1) as {\em external} vertices, and the two inner vertices (that have degree~2) as {\em internal} vertices.

    For $d > 1$, we replace every vertex of $4P_1$ by a cluster of $d$ vertices. The clusters corresponding to external (internal, respectively) vertices of the path are referred to as external (internal, respectively) clusters. Within every external cluster there are no edges (it is an independent set), and within every internal cluster all edges are present (it is a clique). Between two clusters whose corresponding path vertices share an edge we have a complete $d$ by $d$ bipartite graph.  Between two clusters whose corresponding path vertices do not share an edge we have no edges (an empty $d$ by $d$ bipartite graph). Every vertex in $4P_d$ is both part of a clique of size $d+1$ (it has all edges to one of the internal clusters, and that cluster is a clique of size $d$) and part of an independent set of size $d+1$ (it has no edges to one of the external clusters, and that cluster is an independent set of size $d$).  
\end{proof}

Our main existential result is an upper bound on $k(n)$ that matches the lower bound of Proposition~\ref{pro:lowerBound}, up to low order terms. To state an exact expression for the upper bound, it is convenient to consider the function $n(k)$, which is the inverse of the function $k(n)$. Namely, $n(k)$ is the smallest value of $n$ such that there is a $k$-enabling graph on $n$ vertices. Instead of stating an upper bound on $k(n)$, we state a lower bound on $n(k)$.


\begin{restatable}[]{theorem}{tightupperbound}
For all $m \ge 2$ and $k \ge m^3$, $n(k) \geqslant (4 - \frac{5}{m}) k$.
\label{tightupperbound}
\end{restatable}

In Theorem~\ref{tightupperbound}, $m$ can be chosen as $k^{1/3}$ (rounded down to nearest integer), which can easily be seen to imply  that $k(n) \le  \frac{n}{4} + O(n^{2/3})$. 

The lower and upper bounds that we have for $k(n)$ (Proposition~\ref{pro:lowerBound} and Theorem~\ref{tightupperbound}) nearly match, but still differ in low order terms. A plausible conjecture is that the lower bound is tight.

\begin{conjecture}
    For all $n > 0$, $k(n) = \floor{\frac{n}{4}} + 1$.
\end{conjecture}

Empirically we find that the conjecture holds for $n \le 12$. Thus, counter-examples to the conjecture, if there are any, require graphs with more than 12 vertices.


Another theme of our work concerns computational complexity issues related to $k(n)$. Observe that for a vertex $v$, being $k$-enabling is an NP property (a clique of size $k$ and an independent set of size $k$, both containing $v$, certify that $v$ is $k$-enabling), and so being $k$-excluding is a co-NP property. For $k > \frac{n+1}{2}$, $v$ is necessarily $k$-excluding. It is easy to show that deciding whether a vertex is $k$-enabling is NP-hard, and this holds for every $\varepsilon > 0$ and all $k$ in the range  $n^{\varepsilon} \le k \le \frac{n}{2} - n^{\varepsilon}$. Moreover, for this range of values of $k$, it is also NP-hard to decide whether a graph has a $k$-enabling vertex. (This is proved by a simple reduction from the NP-hard question of whether a graph an $t$ vertices has a $k$ clique, with  $t^{\varepsilon} \le k \le t - t^{\varepsilon}$. Simply append to the graph $k$ isolated vertices.) 

Deciding whether a specific vertex is $k$-excluding is computational equivalent to deciding whether a vertex is $k$-enabling (just negate the answer). Our focus is on deciding whether a graph has a $k$-excluding vertex, and we find that its complexity differs from that of deciding whether a graph has a $k$-enabling vertex.

Theorem~\ref{tightupperbound} implies that for every $\delta > 0$ and sufficiently large $n$, every graph on $n \le (4 - \delta)k$ vertices must have a $k$-excluding vertex. Such a vertex can be found by scanning all vertices, and checking for each vertex whether its set of neighboring vertices fails to have a clique of size $k-1$, or its set of non-neighboring vertices fails to have an independent set of size $k-1$. However, such a procedure might not be efficient, because finding maximum size cliques (and independent sets) is NP-hard. Is there an alternative procedure that runs in polynomial time? Our main complexity results gives a positive answer to this question.


\begin{restatable}[]{theorem}{findexcludingvertex}
    For every fixed $0 < \delta < 1$ and $n \le (4 - \delta)k$, 
    there is a polynomial-time (specifically, $O(n^{1/\delta^2})$) algorithm that finds a $k$-excluding vertex $v$ in graphs with $n$ vertices. (By Theorem~\ref{tightupperbound}, a $k$-excluding vertex must exist, unless $k$ is bounded by some constant that only depends on $\delta$. In this latter case, determining whether the graph is $k$-excluding, and finding a $k$-excluding vertex if there is one, can be done in constant time.) Moreover, the algorithm specifies a requirement (being part of a $k$-clique, or part of a $k$-independent-set) that $v$ fails to satisfy.
    \label{findexcludingvertex}
\end{restatable}

Complementing \cref{findexcludingvertex}, we present an NP-hardness result that shows that the relation between $n$ and $k$ in \cref{findexcludingvertex} is nearly best possible.

\begin{restatable}[]{proposition}{nphardness}
For every $\varepsilon > 0$ and $n \ge (4 +\varepsilon)k$, deciding whether a graph on $n$ vertices is $k$-enabling is NP-hard.
\label{nphardness}
\end{restatable}

\subsection{Related work}


The questions studied in this work are related to a fundamental cryptographic primitive known as {\em secret sharing}~\cite{secretsharing,shamir}. 
In its simplest form, $(2, n)$-threshold scheme for $1$ bit, it boils down to the following problem: $1$ bit of information $S$ should be shared between $n$ participants. Each participant $i$ receives a share (bit string) $S_i$ that by itself cannot be used in order to deduce $S$. However, every two participants $i$ and $j$, $i \neq j$, can use $S_i$ and $S_j$ to efficiently and reliably find the value of $S$.

One of the possible secret sharing schemes goes as follows: for $n$ participants, one creates a graph $G$ consisting of $n$ disjoint components, where each component is a copy of $K_n$ (the full graph on $n$ vertices). The name of each vertex has two ``coordinates": the component's index and the vertex's index inside the component. Participant $i$ receives a share $S_i$ composed of two parts. One is the name $k$ of a component, and needs to be given explicitly to $i$. The other is a vertex within the component, and this vertex need not be given explicitly, as by convention, it is the $i$th vertex in the component. Hence, the share $S_i = (k,i)$ is a vertex in the graph $G$. To deduce the secret $S$, two participants $i$ and $j$ check if their respective vertices $S_i$ and $S_j$ share an edge in $G$. If yes, $S = 1$, otherwise $S=0$. The case for $n = 4$ is depicted in the pictures below. Selected vertices are shown in red. If $S = 1$, all vertices are picked from the same component ($k$ is picked uniformly at random, and the same $k$ is used for all shares) and form a clique. If $S = 0$, all vertices are picked from different components (the respective values of $k$ used for the shares $(S_1,S_2,S_3,S_4)$ are arranged in cyclic order, starting at a uniformly random $k$) and form an independent set.

\begin{figure}[ht]
    \centering
    \includegraphics[scale=0.3]{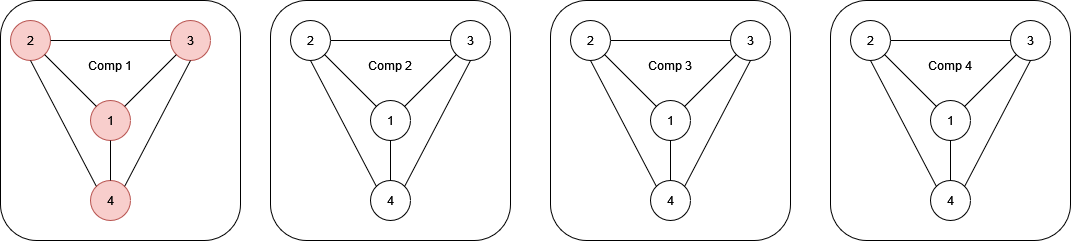}
    \caption{Clique-IS secret sharing scheme for $n = 4$. One realization for the share value $S = 1$ is depicted. There are four possible realizations, differing only in the component chosen.}
    \label{4K4clique}
\end{figure}
\begin{figure}[ht]
    \centering
    \includegraphics[scale=0.3]{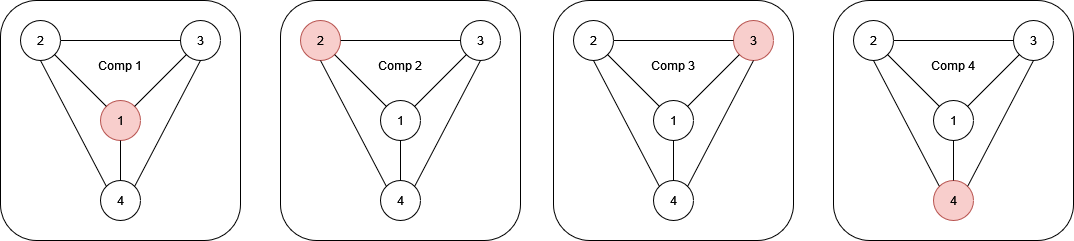}
    \caption{Clique-IS secret sharing scheme for $n = 4$. One realization for the share value $S = 0$ is depicted. There are four possible realizations, by taking cyclic shifts of the names of the components.}
    \label{4K4IS}
\end{figure}

This scheme is information-theoretically secure: the distribution of $S_i$ when $S=0$ is identical to the distribution of $S_i$ when $S=1$.  





Every possible $(2, n)$-threshold scheme for $1$ bit can be represented as some graph $G$ in which cliques encode $S=1$ and independent sets encode $S=0$. Suppose that the size of each participant's share is no more than $k$ bits, and let $K = 2^k$. Create a graph $G$ on $nK$ vertices. Each vertex has $2$ indices, $(i, S_i)$, where $i$ represents a participant's name, and $S_i$ represents a share value.
Two vertices $(i, S_i)$ and $(j, S_j)$ are connected by an edge if and only if participant $i$ with share $S_i$ and participant $j$ with share $S_j$ will deduce together that the secret $S$ has value~1. For pairs of vertices that cannot arise as pairs of shares (for example, if $i=j$), edges can be put arbitrarily (they have no effect on the secret sharing scheme). 

Suppose that a graph $G$ represents a $(2,n)$ secret sharing scheme. Then every vertex $v$ in $G$ that might be chosen as a share must be both a member of a clique of size  $n$ and an independent set of size $n$. Otherwise, the agent that receives $v$ can deduce at least one value that $S$ cannot have. Consequently, every vertex of $G$ that can be used as a share is $n$-enabling, in our terminology. Moreover, every vertex that is $n$-excluding can be removed from $G$ without affecting the secret sharing scheme. Consequently, in {\em minimal} graphs that represent $(2,n)$ secret sharing schemes, all vertices are $n$-enabling. Renaming $n$ as $k$, this motivates the existential part of our study concerning how small can graphs be and still have all their vertices $k$-enabling.



Having all vertices $n$-enabling is a necessary condition for $(2,n)$ secret sharing schemes, but not a sufficient condition. This property by itself only implies that upon receiving a share, a participant does not know for sure the value of $S$. However, it does not exclude the possibility that a participant can have quite a good guess of what the value of $S$ is (e.g., conditioned on receiving a certain share value $S_i$, it might be that $S=1$ is much more likely than $S=0$). In fact, if one wishes to prevent any information from leaking about $S$, the graph presented above, composed of $n$ disjoint cliques, gives the smallest possible shares. (The size of the share that needs to be communicated to a participant is $\log n$ bits, because for each participant, only $n$ out of the $n^2$ vertices of the graph can serve as her share.  Kilian and Nisan (in an unpublished manuscript, see for example~\cite{BogdanovGK20}) showed that $\log n$ is the smallest possible share size for $(2,n)$ threshold secret sharing schemes.)  

In at attempt to achieve secret sharing schemes with smaller share values, it was suggested in~\cite{secretsharing} to replace information theoretical security by computational security. Under this definition, the share value $S_i$ might be correlated with the value of $S$, but making use of this correlation is computationally intractable. Under such a relaxed definition of security, the argument why $G$ should not have $k$-excluding vertices breaks down. Due the the NP-hardness of finding cliques and independent sets, it may be okay if a vertex that is $n$-excluding is used as a share value, as long as it is intractable to determine that it is $n$-excluding, or what makes it $n$-excluding (is it because it is not part on a clique of size $n$, or because it is not part on an independent set of size $n$). This motivates the complexity part of our study, concerning whether it is tractable to find $k$-excluding vertices. 

We now mention some other related work.

A relaxed version of $k$-enabling and $k$-excluding vertices is one in which the requirement of being part of a $k$-clique is dropped. That is, given an $n$ vertex graph, we seek the largest $k$ such that every vertex is part of an independent set of size $k$. In this case, $k$ can be as large as $n$ (if the graph has no edges). From a complexity point of view, determining the largest value of $k$ is NP-hard. This is a consequence of known hardness results for computing the chromatic number. We state one such result which has a relatively simple proof (see Section 4.1 in~\cite{FeigeK98}). 

\begin{theorem}
    \label{thm:FK98}
    For some constant $q > 0$, it is NP-hard to distinguish between graphs on $n$ vertices in which every vertex is in an independent set of size $\frac{n}{6}$, and graphs in which no vertex is in an independent set larger than $(1 - q)\frac{n}{6}$.
\end{theorem}

We remark that much larger multiplicative gaps (of the form $n^{1 - \epsilon}$) in the approximability were proved under randomized reductions in~\cite{FeigeK98}, and under deterministic reductions in~\cite{Zuckerman07}.


A classic question in Ramsey theory asks how {\em large} should $n$ be so that {\em every} $ n$-vertex graph either has a clique of size $k$ or an independent set of size $k$. The answer is known to lie between $2^{k/2}$ and $2^{2k}$ \cite{erdos}. One of the questions asked in this paper is how {\em small} should $n$ be so that {\em at least one} $ n$-vertex graph has the property that {\em every} vertex is part of both a clique of size $k$ and an independent set of size $k$. We show that the answer is roughly $4k$.


\section{Upper bound on \texorpdfstring{$k(n)$}{k(n)}}


       We say that two sets $X$ and $Y$ are \textbf{almost disjoint} if $|X \cap Y| \le 1$.
Clearly, in any graph, every clique and independent set are almost disjoint. 

In our proof of Theorem~\ref{tightupperbound}, we shall use the following definition.

\begin{definition}
    A clique-IS system of graph $G$ of order $m$, or $m$-system for short, is a collection of $m$ independent sets $I_1 \dots I_m$ and $m$ cliques $C_1 \dots C_m$ such that all $I_i$ are disjoint and all $C_i$ are disjoint. The size of an $m$-system is the number of vertices contained in the union of all $I_i$ and all $C_i$. 
\end{definition}

\begin{observation}
    \label{observ}
    If graph $G(V,E)$ contains the $m$-system of size $L$, then $|V| \ge L$.
\end{observation}

\begin{claim}
    \label{systemsize}
    The size of an $m$-system is no less than $\sum_{i=1}^{m} |I_i| + \sum_{i=1}^{m} |C_i| - m^2$.
\end{claim}

\begin{proof}
    Follows easily from the fact that the independent sets are disjoint, the cliques are disjoint, and a pair of clique and independent set intersect in at most one vertex.    
\end{proof}

As an illustration of how to use $m$-systems towards the proof of Theorem~\ref{tightupperbound}, we start with the simplest case of $m=1$.


\begin{proposition}
    \label{2k-1}
    $n(k) \geqslant 2k-1$.
\end{proposition}

\begin{proof}
    Consider an arbitrary $k$-enabling graph. Pick an arbitrary vertex $v$. Being $k$-enabling, $v$ is contained in an independent set $I_1$ of size at least $k$ and in a clique $C_1$ of size at least $k$. This gives a 1-system $I_1, C_1$, and by Claim~\ref{systemsize} its size (which we denote by $L_1(n,k)$) is at least $|I_1| + |C_1| - 1^2 \ge 2k-1$. By \cref{observ}, $n \ge L_1(n, k) \ge 2k - 1$.
\end{proof}

To further illustrate our techniques, we now consider $m$-systems for $m=2$.

\begin{claim}
\label{2-system}
    Every $k$-enabling graph contains a $2$-system with $|I_1|, |C_1| \geqslant k$ and $|I_2|, |C_2| \geqslant k_2 = \frac{k (k - 1)}{n - k}$.
\end{claim}

\begin{proof}
Consider an arbitrary $k$-enabling graph $G(V,E)$.
    Pick an arbitrary vertex $v$. Vertex $v$ is both in a clique $C$ of size $k$ and an independent set $I$ of size $k$. 
    
    Each vertex $w$ of $I$ is in a clique $C_w$ of size $k$. 
    $I$ and $C_w$ are almost disjoint. Therefore, each vertex in $I$ has $k - 1$ edges going out of $I$. In total, there are $k \times (k - 1)$ such edges. Their other end-points cannot be in $I$ (because $I$ is an independent set), so they are in $V \setminus I$. $|V \setminus I| = n - k$, so the average $I$-degree of vertices in $V \setminus I$ is $k_2 = \frac{k (k - 1)}{n - k}$, and there is at least one vertex $u_2$ with $I$-degree at least $k_2$.

    Vertex $u_2$ is a part of an independent set $I_{u_2}$ of size $k$. No more than $|I| - k_2 = k - k_2$ vertices in $I$ can be in $I_{u_2}$, because at least $k_2$ vertices of $I$ are connected to $u_2$. So $I_{u_2}$ contains $u_2$, at most $k - k_2$ vertices in $I$ and at least $k_2 - 1$ other vertices. 
    The union of $u_2$ and these other vertices are a subset of $I_{u_2}$, and hence they form an independent set $I_2$ of size at least $k_2$ which is disjoint from $I$. 
    
    A similar reasoning, replacing edges with non-edges and vice versa, provides a clique $C_2$ of size at least $k_2$ that is disjoint from $C$.
\end{proof}   
    
    \begin{corollary}
        \label{3k}
        $n(k) \ge 3k - 3$ for $k \ge 3$.
    \end{corollary}
    \begin{proof}
        By \cref{2-system}, every $k$-enabling graph contains a $2$-system of size $L_2(n, k) \geqslant |I_1| + |C_1| + |I_2| + |C_2| - 2^2 \geqslant 2k + 2\frac{k(k - 1)}{n - k} - 4$. By \cref{observ}, $n \geqslant L_2(n, k) \geqslant 2k + 2\frac{k(k - 1)}{n - k} - 4$. The rest of the proof follows by simple algebraic manipulations that we detail here for completeness.
        
    The requirement $n \geqslant 2k + 2\frac{k(k - 1)}{n - k} - 4$ is equivalent to the requirement that $n - 2k - 2\frac{k(k - 1)}{n - k} + 4 \geqslant 0$.
    $$n - 2k - 2\frac{k(k - 1)}{n - k} + 4 = \frac{(n - 2k)(n - k) - 2k(k-1) + 4(n - k)}{n-k}$$
    Proposition~\ref{2k-1} implies that the denominator is positive. Hence, the numerator is required to be positive. The numerator simplifies to:
    $$(n - 2k)(n - k) - 2k(k-1) + 4(n - k) = n^2 + (4-3k)n - 2k$$
    For $n = 3k - 4$. $n^2 + (4-3k)n - 2k = -2k$, which is negative for all positive $k$. Hence, $n \ge 3k - 3$. 
    \end{proof}

We now consider $m$-systems for general $m$.

\begin{lemma}
\label{superlemma}
    For given $n$ and $k$, let $k_2 = \frac{k (k - 1)}{n - k}$ and for $j \geqslant 2$, $k_{j + 1} = \frac{(k + k_j)(k - j)}{n - k - k_j}$. 
     Then, for every $m \ge 3$, every $k$-enabling graph $G$ on $n$ vertices contains an $m$-system with $|I_1|, |C_1| \ge k$ and $|\bigcup\limits_{1 \leqslant i \leqslant j} I_i| \ge k + k_j$ for every $j \ge 2$ (and likewise, $|\bigcup\limits_{1 \leqslant i \leqslant j} C_i| \ge k + k_j$). In the $m$ system, we allow some of the sets $I_i$ or $C_i$ to be empty (this is unavoidable, as the lemma allows $m > n$), and note that it is not claimed that the sequence $k_j$ as defined is non-decreasing.
\end{lemma}

\begin{proof}    
    For every fixed $m$, the proof is by induction on {$j$}.
    
    The case {$j = 2$} serves as the base case of the induction and was proved in \cref{2-system}.

    We now assume that the lemma holds for $j$ and prove it for $j+1$. 
    By the induction assumption the $k$-enabling graph $G(V,E)$ has an $j$-system: disjoint cliques {$C, C_2, \dots, C_j$} and disjoint independent sets {$I, I_2, \dots I_j$}, each with total sizes $k + k_j$. 
    Let $\hat{I}$ denote $\bigcup\limits_{1 \leqslant i \leqslant j} I_i$.
    Each vertex $v \in \hat{I}$  
    is a part of clique of size at least $k$. At most {$j$}  vertices of this clique {are} in $\hat{I}$ 
    (no more than one vertex in each of the independent sets), so there are at least {$k - j$} edges joining $v$ to vertices in {$V \setminus \hat{I}$}. There are in total at least {$(k + k_j)(k - j)$} edges between $\hat{I}$ and $V \setminus \hat{I}$. 

    The average {$\hat{I}$}-degree of a vertex in {$V \setminus \hat{I}$} is at least {$k_{j + 1} = \frac{(k + k_j)(k - j)}{n - (k + k_j)}$}. Therefore, there is at least one vertex {$u_{j + 1}$} with {$\hat{I}$}-degree of at least {$k_{j + 1}$}. {$u_{j + 1}$} is part of an independent set {$I_{u_{j + 1}}$} of size $k$. Let {$I_{j + 1} = I_{u_{j + 1}} \setminus \hat{I}$}. (Note that $I_{j + 1}$ might be empty.) The $|I_{j+1}| + |\hat{I}|$ is at least $k$ plus the number of neighbors of $u_{j+1}$ in $\hat{I}$ (as none of these neighbors is in $I_{j+1}$), giving the desired lower bound of $k + k_{j+1}$.

    Similar reasoning, replacing edges with non-edges and vice versa, provides a set {$C_{j + 1}$} with $|\bigcup\limits_{1 \leqslant i \leqslant j+1} C_i| \ge k + k_{j+1}$.
    
    Finally, note that the $k_j$ values need not be integers. Thus, they need to be rounded up to the nearest integer. This does not affect our proofs, because increasing $k_j$ only increases the expression $\frac{(k + k_j)(k - j)}{n - k - k_j}$ for $k_{j+1}$. 
    \end{proof}

\begin{corollary}
    \label{kkm}
    For every $m \ge3$, in every $k$-enabling graph, $n$ needs to satisfy $n \geqslant 2(k + k_m) - m^2$, for $k_m$ as defined in Lemma~\ref{superlemma}.
\end{corollary}

\begin{proof}
    Let $G$ be a $k$-enabling graph $G$ on $n$ vertices. By \cref{superlemma}, $G$ contains an $m$-system with $|\bigcup\limits_{1 \leqslant i \leqslant m} I_i| \ge k + k_m$ and $|\bigcup\limits_{1 \leqslant i \leqslant m} C_i| \ge k + k_m$. By Claim~\ref{systemsize} its size is at least $2(k + k_m) - m^2$, and by \cref{observ}, $n$ is at least as large.    
\end{proof}

It remains to provide lower bounds on $k_m$. 

\begin{lemma}
    \label{lem:km}
    For $m \ge 2$ and $n \le 4k - 3m$ it holds that $k_j \ge (1 - \frac{2}{j+1})k$ for every $2 \le j \le m$.
\end{lemma}

\begin{proof}
For fixed $m$, we prove the lemma by induction on $j$.

The base case is $j=2$. We have from Claim~\ref{2-system} that $k_2 = \frac{k(k-1)}{n-k} \ge \frac{k(k-1)}{3k - 3m} \ge \frac{k}{3}$, as desired.

For the induction step, we assume that  $k_j \ge (1 - \frac{2}{j+1})k$ and need to show that $k_{j+1} \ge (1 - \frac{2}{j+2})k$.  
Using the expression provided for $k_{j+1}$ in Lemma~\ref{superlemma}, we have that:

$$k_{j + 1} = \frac{(k + k_j)(k - j)}{n - k - k_j} \ge \frac{(k + k_j)(k - m)}{4k - 3m - k - k_j} \ge \frac{(k + k_j)k}{3k - k_j} \ge \frac{2 - \frac{2}{j+1}}{2 + \frac{2}{j+1}} \cdot k = (1 - \frac{2}{j+2})k$$
\end{proof}

We now restate and prove Theorem~\ref{tightupperbound}.

\tightupperbound*
\begin{proof}
Observe that the parameters of the theorem are such that $\frac{5k}{m} \ge 3m$, so $n$ is in the range in which Lemma~\ref{lem:km} applies. By \cref{kkm} and Lemma~\ref{lem:km}, $n(k) \ge 2(k + k_m) -m^2 \geqslant 2(2 - \frac{2}{m+1})k - m^2 \geqslant 4k - \frac{4k}{m+1} - m^2$.  As $k \ge m^3$, we have that $m^2 \le \frac{k}{m}$, implying that $n(k) \ge  (4 - \frac{5}{m}) k$. 
\end{proof}

\section{An algorithm for finding  \texorpdfstring{$k$}{k}-excluding vertices}

In this section we prove Theorem~\ref{findexcludingvertex}. We shall use terminology that is introduced in the following definition.

\begin{definition}
\label{eps-m-system}
 For $0 < \varepsilon < 1$, we introduce the following notions:\\
    $\varepsilon$-almost-clique $C$ is a subgraph in which each vertex of $C$ has $C$-degree at least $(1-\varepsilon)|C|$.\\
    $\varepsilon$-almost-IS $I$ is a subgraph in which each vertex of $I$ has $I$-degree at most $\varepsilon |I|$. \\
    $\varepsilon$-almost-Clique-IS system of graph $G$ of $m$th order, or $\varepsilon$-$m$-system for short, is a collection of $m$ $\varepsilon$-almost ISs $I_1 \dots I_m$ and $m$ $\varepsilon$-almost-cliques $C_1 \dots C_m$ such that all $I_i$ are disjoint and all $C_i$ are disjoint.\\ The size of an $\varepsilon$-$m$-system is the number of vertices in the union of all $I_i$ and all $C_i$.\\
\end{definition}

\begin{remark}
    For \cref{eps-m-system} to make sense, $\varepsilon|C| \geqslant 1$ must hold. For our choices of $\varepsilon$ and the sets $I$ that we will consider, $\varepsilon|I| \geqslant 1$ will also hold.  
\end{remark}

\begin{observation}
\label{obs:ISinAlmostClique}
    An $\varepsilon$-almost-clique $C$ cannot contain an independent set of size strictly larger than $\varepsilon|C|$. An $\varepsilon$-almost-IS $I$ cannot contain a clique of size strictly larger than $\varepsilon|I| + 1$.
\end{observation}

\begin{proof}
    Let $I'$ be an independent set in $C$. Pick any vertex $v \in I'$. None of $v$'s neighbors is in $I'$, so $|C| \ge |I'| + (1 - \varepsilon)|C|$, implying that $|I'| \le \varepsilon |C|$.
    
    Let $C'$ be a clique in $I$. Pick any vertex $v \in C'$. All other vertices of $C'$ are neighbors of $v$, implying that $|C'| \le \varepsilon |I| + 1$.
\end{proof}

\begin{lemma}
    \label{intersection}
        Let $C_i$ be an $\varepsilon$-almost clique and let $I_j$ be an $\varepsilon$-almost independent set. Then $|C_i \bigcap I_j| \leqslant \varepsilon (|C_i| + |I_j|)$
    \end{lemma}

    \begin{proof} Consider any vertex $v \in C_i \bigcap I_j$.
        On one hand, $deg_{C_i \bigcap I_j}(v) \leqslant deg_{I_j}(v)$. On the other hand, consider $\overline{deg}$, the number of non-edges. $\overline{deg}_{C_i \bigcap I_j}(v) \leqslant \overline{deg}_{C_i}(v)$.
        $deg_{C_i \bigcap I_j}(v) + \overline{deg}_{C_i \bigcap I_j}(v) = |C_i \bigcap I_j| - 1$.
        Thus, $deg_{I_j}(v) + \overline{deg}_{C_i}(v) \geqslant |C_i \bigcap I_j| - 1$.
        By definition, $deg_{I_j}(v) \leqslant \varepsilon|I_j|$ and $\overline{deg}_{C_i}(v) = |C_i| - 1 - {deg}_{C_i}(v) \leqslant \varepsilon|C_i| - 1$. Adding all of this yields: $\varepsilon|I_j| + \varepsilon|C_i| - 1 \geqslant deg_{I_j}(v) + \overline{deg}_{C_i}(v) \geqslant |C_i \bigcap I_j| - 1$, implying $\varepsilon (|C_i| + |I_j|) \geqslant |C_i \bigcap I_j|$.
    \end{proof} 

\begin{claim}
    \label{epsilonsystemsize}
    The size of an $\varepsilon$-$m$-system is at most $(1 - m \varepsilon)\left(\sum_{i=1}^{m} |I_i| + \sum_{i=1}^{m} |C_i|\right)$.
\end{claim}

\begin{proof}
    For all $i, j$, $I_i$ and $I_j$ are disjoint, and $C_i$ and $C_j$ are disjoint. Using this fact, we can lower bound the size of the $\varepsilon$-$m$-system by

    $\sum\limits_{i=1}^{m} |I_i| + \sum\limits_{i=1}^{m} |C_i| -  \sum\limits_{i=1}^{m} \sum\limits_{j=1}^{m} |I_i \bigcap C_j| \geqslant {\sum\limits_{i=1}^{m} |I_i| + \sum\limits_{i=1}^{m} |C_i| - \sum\limits_{i=1}^{m} \sum\limits_{j=1}^{m} \varepsilon (|C_i| + |I_j|) =\ }(1-m\varepsilon)\left(\sum\limits_{i=1}^{m} |I_i| + \sum\limits_{j=1}^{m} |C_j|\right)$

    \noindent where the first inequality follows from Lemma~\ref{intersection}.
\end{proof}

In our proof of Theorem~\ref{findexcludingvertex} we shall use the following Theorem, whose proof is based on a {simple} modification of an algorithm of~\cite{FS97}.

\begin{theorem}
\label{thm:FS}
    For every $\varepsilon \ge \frac{2}{k}$, there is an algorithm running in time $n^{O(g(\varepsilon))}$, where $g(\varepsilon) = \frac{1 + \log{\frac{n}{k}}}{\varepsilon}$ (in our case $k < n < 4k$, implying that $0 < \log{\frac{n}{k}} < \log{4}$, and $g(\varepsilon) < \frac{3}{\varepsilon}$), that on input of a graph $G(V,E)$ (here, $n = |V|$) and integer $k$, returns one of the following {two outputs:

    \begin{enumerate}
        \item Determines that there is no clique of size $k$ in $G$.
        \item Does not determine whether there is a clique of size $k$ or not, but returns an {$\varepsilon$-almost-clique $C$ in $G$ with $|C| \geqslant k$.} In particular, this subgraph does not have an independent set of size larger than $\varepsilon |C|$.
    \end{enumerate}}
\end{theorem}

\begin{proof}
    An $\varepsilon$-almost-clique $C$ in $G$ with $|C| \geqslant k$ will be referred to as an \textbf{acceptable} graph. Consider the following algorithm (a simple modification of algorithm \textbf{DenseSubgraph} from~\cite{FS97}):
    \begin{algorithm}[H]
    \caption{FindAcceptableGraph($G(V,E), k$)}
    \begin{algorithmic}
    {
        \If{$|V| < k$} \Return{$\varnothing$}
        \EndIf
        \State $v \gets$ vertex with smallest degree in $G$
        \State $h \gets deg(v)$
        \If{$h < (1 - \varepsilon)|V|$} 
            \State $G_v \gets$ subgraph of $G$ induced by $v$ and its neighbors
            \State $H \gets FindAcceptableGraph(G_v, k)$
            \If{$H \neq \varnothing$} \Return{$H$}
            \Else\ \Return{$FindAcceptableGraph(G \setminus \{v\}, k)$}
            \EndIf
        \Else\ \Return{$G$} \Comment{$h \geqslant (1 - \varepsilon)|G|$}
        \EndIf
            }

    \end{algorithmic}
    \label{alg_reasonable}
    \end{algorithm}

        If the algorithm returns a non-empty graph, then it is an \textbf{acceptable} graph. This is because
        {recursive calls either return their input graph if this graph is \textbf{acceptable}, or an empty graph.}

        \begin{claim}
        \label{cl:reasonable}
        If $G$ contains a clique of size $k$, then the algorithm returns an \textbf{acceptable} graph.
    \end{claim}

    \begin{proof}
        By induction on the graph size. If $n < k$, there is no clique {of size $k$, and indeed the algorithm} returns an empty graph. {Now consider arbitrary $N \ge k$. We need to prove that the statement holds for $n$, assuming that the induction hypothesis holds for all $n < N$.  We consider} two cases:
        \begin{itemize}
            \item Smallest degree in the graph is at least $(1 - \varepsilon)N$. Then the graph itself is \textbf{acceptable}, {and the algorithm} 
            returns it.
            \item Otherwise, if smallest degree in the graph is less than $(1 - \varepsilon)N$, the algorithm makes two recursive calls for graphs of smaller size. 
            Since by the {inductive} hypothesis the graph contains a clique of size $k$, one of the graphs ($G_v$ and $G_{-v}$) should contain it, depending on whether the vertex with the smallest degree is a part of it. Therefore, by induction hypothesis applied to both of them and the fact that the algorithm returns either an \textbf{acceptable} or an empty graph (and for one of them it would be non-empty), it follows that the output would also be \textbf{acceptable}.
        \end{itemize}     
    \end{proof}

    \begin{claim}
    \label{cl:time}
        The algorithm works in $n^{O(\frac{1 + \log{\frac{n}{k}}}{\varepsilon}))}$ time.
    \end{claim}
    \begin{proof}
        A single iteration of the algorithm spends polynomial time (actually, no more than quadratic) and makes two recursive calls: one of size $n - 1$ and one of size at most $(1 - \varepsilon) n + 1$. Denoting the running time by $T(n)$, then $T(n) \leqslant O(n^c) + T(n - 1) + T((1 - \varepsilon) n + 1)$. As shown in \cite{FS97} solution to this recurrence is $n^{O(\frac{1 + \log{\frac{n}{k}}}{\varepsilon}))}$.
    \end{proof}

    The combination of Claims~\ref{cl:reasonable} and~\ref{cl:time} prove Theorem~\ref{thm:FS}.   
\end{proof}

We now restate and prove Theorem~\ref{findexcludingvertex}.

\findexcludingvertex*

\begin{proof}
Fix $\delta > 0$. Suppose that $n \le (4-\delta)k$. We show an algorithm that finds a $k$-excluding vertex. The running time of the algorithm will be  $n^{O(\delta^2)}$. We shall assume that $k > K_{\delta}$ for some $K_{\delta}$ that depends only on $\delta$.

We shall construct a system of disjoint {$\varepsilon$-almost-cliques}, $C_1, C_2, \ldots$, and disjoint {$\varepsilon$-almost-independent sets} $I_1, I_2, \ldots$ . 
For every $j \le 1$, we use $c_j = \sum_{i=1}^j |C_i|$ to denote the sum of sizes of the first $j$ {$\varepsilon$-almost-cliques}. 
We describe the construction of the {$\varepsilon$-almost-cliques}. The construction for {$\varepsilon$-almost-IS} is similar (and in fact, slightly simpler, due to the asymmetry of cliques and independent sets in Observation~\ref{obs:ISinAlmostClique}). 

{Use the algorithm of Theorem~\ref{thm:FS} to find $C_1$, an $\varepsilon$-almost clique of size at least $k$ in $G$.} Thereafter, for every $j$, we construct $C_{j+1}$ as follows.
By Observation~\ref{obs:ISinAlmostClique}, the largest independent set in $\bigcup_{i=1}^j C_i$ is of size at most $\varepsilon c_j + j$. This implies that every $k$-enabling vertex in $\bigcup_{i=1}^j C_i$ has at least $k - \varepsilon c_j - j - 1$ non-edges to vertices outside $\bigcup_{i=1}^j C_i$. If a vertex $u \in \bigcup_{i=1}^j C_i$ fails this condition, $u$ is declared as $k$-excluding (and moreover, the reason for this declaration is determined to be that $u$ is not part of a $k$-independent-set), and the algorithm ends. 
If no vertex in $\bigcup_{i=1}^j C_i$ is declared as $k$-excluding, then there is a vertex $v \not\in \bigcup_{i=1}^j C_i$ with at least $\frac{c_j(k - \varepsilon c_j - j - 1)}{n - c_j}$ non-edges to $\bigcup_{i=1}^j C_i$. Hence, 
at least $k - (c_j - \frac{c_j(k - \varepsilon c_j - j - 1)}{n - c_j})$ {vertices} of the $k$-clique of $v$ are outside {of} $\bigcup_{i=1}^j C_i$ {and form a clique. Denote this lower bound on its size by $n_{j + 1}$.}

{Apply the algorithm of \cref{thm:FS} to the graph which includes $v$ and those of its neighbors which do not belong to $\bigcup_{i=1}^j C_i$, with size of potential clique (the parameter $k$ in \cref{thm:FS}) equal to {$n_{j+1}$}. We either find a new $\varepsilon$-almost-clique {$C_{j+1}$} of size at least $n_{j+1}$, or else, deduce that $v$ is not part of $k$-clique in $G$, and hence is $k$-excluding.} 
In the former case, we have that $c_{j+1} \ge c_j + n_j = k + \frac{c_j(k - \varepsilon c_j - j - 1)}{n - c_j}$.

Fixing some value $m \ge2$, by repeating the above process we construct a system of $m$ disjoint {$\varepsilon$-almost-cliques} (some of which might be empty) of total size at least $c_m$. In analogy to Lemma~\ref{lem:km}, we have the following lemma.

\begin{lemma}
    \label{lem:cm}
    For $m \ge 2$ and $n \le 4k - 6\varepsilon k - 3(m+1)$ it holds that $c_j \ge (2 - \frac{2}{j+1})k$ for every $2 \le j \le m$.
\end{lemma}

\begin{proof}
For fixed $m$, we prove the lemma by induction on $j$. 

The base case is $j=2$. As $c_1 = k$, we have that $c_2 \ge k + \frac{k(k - \varepsilon k - 2)}{n - k} \ge k + \frac{k(k - \varepsilon k - 2)}{3k - 6\varepsilon k - 3(m+1)} \ge \frac{4k}{3}$, as desired.

For the induction step, we assume that  $c_j \ge (2 - \frac{2}{j+1})k$ and need to show that $c_{j+1} \ge (2 - \frac{2}{j+2})k$, or equivalently, that $c_{j+1} - k \ge (1 - \frac{2}{j+2})k$. Note that in the range $c_j \le 2k$ it holds that $\varepsilon c_j \le 2\varepsilon k$.
We have that:

$$c_{j + 1} - k \ge \frac{c_j(k - \varepsilon c_j - j - 1)}{n - c_j} \ge \frac{c_j(k - 2\varepsilon k - m - 1)}{3k - 6\varepsilon k - 3(m+1) - (c_j - k)} \ge \frac{c_j \cdot k}{3k - (c_j - k)} \ge \frac{2 - \frac{2}{j+1}}{2 + \frac{2}{j+1}} \cdot k = (1 - \frac{2}{j+2})k$$
\end{proof}

{
Arguments analogous to those above provide a collection of $m$ disjoint $\varepsilon$-almost-IS of total size at least {$(1 - \frac{2}{m+1})k$}. By \cref{epsilonsystemsize} the total size of this $\varepsilon$-$m$-system is at least {$2(1 - \varepsilon m)(2 - \frac{2}{m+1})k$}.
}
If the algorithm fails to find a {$k$-excluding} vertex, we get that {$(4 - \delta)k \ge 2(1 - \varepsilon m)(2 - \frac{2}{m+1})k$}, implying that $\delta \le 4(\epsilon m + \frac{1}{m+1}$). Choosing, $\varepsilon = \frac{1}{m(m+1)}$ we get that if $\delta > \frac{8}{m+1}$ the algorithm finds a $k$-excluding vertex. Note that for Lemma~\ref{lem:cm} to apply we need that $\delta k \ge 6\epsilon k + 3(m+1)$,
which holds if $k > (m+1)^2$.

Summarizing, for every constant $\delta > 0$, set $m$ to be the largest integer smaller than $\frac{8}{\delta} - 1$, set $\epsilon = \frac{1}{m(m+1)}$ and set $K_{\delta} = {(m+1)^2}$ (note that $K_{\delta}$ is a constant that only depends on $\delta$). For these parameters, for every $k > K_{\delta}$ and every input graph on $n \le (4 - \delta)k$ vertices, the algorithm necessarily finds a $k$-excluded vertex.

The above algorithm makes $m$ 
calls to the {\textbf{FindAcceptableGraph} algorithm}, which works in $n^{O(\frac{1}{\varepsilon})}$ time. Since {$\varepsilon = O(\delta^2)$}, 
the running time is 
$n^{O(1/\delta^2)}$.
\end{proof}

\section{NP-hardness}

Here we restate and prove Proposition~\ref{nphardness}.

\nphardness*

\begin{proof}
    Let $G_1$ be a graph of $\varepsilon k$ vertices in which we wish to determine if every vertex is in an independent set of size $\frac{\varepsilon k}{6}$ (a {\em yes} instance), or some vertex is not in an independent set of size $\frac{\varepsilon k}{6}$ (a {\em no instance}). By Theorem~\ref{thm:FK98}, this is an NP-hard problem. We reduce it to the problem of deciding whether a graph $G$ is $k$-enabling, with parameters as stated in Proposition~\ref{nphardness}.

    Set $d = k$ and recall the $(d+1)$-enabling graph $4P_d$ from Proposition~\ref{pro:lowerBound}. Construct the graph $G$ whose set of vertices is the union of those of $4P_k$ and $G_1$, and whose set of edges includes all edges of $4P_k$, all edges of $G_1$, and a complete bipartite graph between the vertices of $G_1$ and a set $S$ of $3k + \frac{\varepsilon k}{6}$ vertices from $4P_k$, where the set $T$ of excluded vertices (those not $S$) is composed of  $k -\frac{\varepsilon k}{6}$ vertices that all belong to one of the external clusters of $4P_k$.

    The graph $G$ contains $(4 + \varepsilon)k$ vertices, as desired. Every vertex in $4P_k$ is $(k+1)$-enabling. Every vertex in $G_1$ is part of a clique of size $k+1$ (because it is fully connected to a clique cluster of $4P_k$). For every vertex $v$ in $G_1$, the largest independent set in $G$ containing it is composed of the union of $T$ and the largest independent set in $G_1$ containing $v$. In {\em yes} instances of $G_1$, its size is at least $k -\frac{\varepsilon k}{6} + \frac{\varepsilon k}{6} = k$, and $G$ is $k$-enabling. In {\em no} instances of $G_1$, its size for some vertex is at most $k -\frac{\varepsilon k}{6} + \frac{\varepsilon k}{6} - 1 = k-1$, and $G$ is not $k$-enabling.
\end{proof}

\subsection*{Acknowledgements}

This research was supported in part by the Israel Science Foundation (grant No. 1122/22). We thank Robert Krauthgamer, Shachar Meir and David Peleg for useful discussions.

\bibliographystyle{plain}

\end{document}